\documentclass[apj]{emulateapj}
\usepackage{amsmath}
\usepackage{epsfig,natbib}
\usepackage{amsmath,amsfonts,amssymb}
\usepackage{mathptmx}
\usepackage{color}
\usepackage{graphicx}
\usepackage{bm}
\slugcomment{}

\begin{document}

\title{Oscillation Driven Magnetospheric Activity In Pulsars}

%\author{
%    Meng-Xiang Lin$^1$, Ren-Xin Xu$^{1,2}$ and Bing Zhang$^{3,1,2}$\\
%    \small $^1$Department of Astronomy, School of Physics, Peking University, \\
%        \small Beijing 100871, P. R. China. \\
%        \small Email: linmx97@gmail.com(MXL); r.x.xu@pku.edu.cn(RXX)\\
%    \small $^2$Kavli Institute for Astronomy and Astrophysics, Peking University, \\
%        \small Beijing 100871, P. R. China \\
%    \small $^3$Department of Physics and Astronomy, University of Nevada, Las Vegas. \\
%        \small Email: zhang@physics.unlv.edu(BZ)
%    }
    
\author{Meng-Xiang Lin\altaffilmark{1}, Ren-Xin Xu\altaffilmark{1,2} and Bing Zhang\altaffilmark{3,1,2,}}

\altaffiltext{1}{Department of Astronomy, School of Physics, Peking University, Beijing 100871, China. Email: linmx97@gmail.com(MXL); r.x.xu@pku.edu.cn(RXX)}
\altaffiltext{2}{Kavli Institute for Astronomy and Astrophysics, Peking University, Beijing 100871, China.}
\altaffiltext{3}{Department of Physics and Astronomy, University of Nevada, Las Vegas, NV 89154, USA. Email: zhang@physics.unlv.edu(BZ)}

%{Submitted to ApJ on Aug. 15th, 2014. Version addressing the referee's comments.}

\begin{abstract}
We study the magnetospheric activity in the polar cap region of pulsars under stellar oscillations. The toroidal oscillation of the star propagates into the magnetosphere, which provides additional voltage due to unipolar induction, changes Goldreich-Julian charge density from the traditional value due to rotation, and hence, influences particle acceleration. We present a general solution of the effect of oscillations within the framework of the inner vacuum gap model, and consider three different inner gap modes controlled by curvature radiation, inverse Compton scattering, and two photon annihilation, respectively. With different pulsar parameters and oscillation amplitudes, one of three modes would play a dominant role in defining the gap properties. When the amplitude of oscillation exceeds a critical value, mode changing would occur. Oscillations also lead to change of the size of the polar cap. As applications, we show the inner gap properties under oscillations in both normal pulsars and anomalous X-ray pulsars / soft gamma-ray repeaters (AXPs/SGRs). We interpret the onset of radio emission after glitches/flares in AXPs/SGRs as due to oscilation-driven magnetic activities in these objects, within the framework of both the magnetar model and the solid quark star model. Within the magnetar model, radio activation may be caused by the enlargement of the effective polar cap angle and the radio emission beam due to oscillation; whereas within the solid quark star angle, it may be caused by activation of the pulsar inner gap from below the radio emission death line due to an oscillation-induced voltage enhancement.  The model can also explain the glitch-induced radio profile change observed in PSR J1119-6127.
\end{abstract}

\section{INTRODUCTION}
Glitches (sudden ``spin-up'' of the star) are detected commonly in pulsars, including both normal pulsars and anomalous X-ray pulsars / soft gamma-ray repeaters (AXPs/SGRs). Stellar oscillations are likely excited during a glitch, which would propagate into the magnetosphere to affect the properties of the inner gap of the pulsar. \citet{Morozova2010, Zanotti2012} have studied the oscillations effect on the magnetospheric activity within the framework of the space-charge-limited flow (SCLF) model.
On the other hand, drifting sub-pulses in pulsar radio emission suggests that there are sub-beams (sparks) generated in the pulsar polar cap region, which can be naturally interpreted as the existence of a inner vacuum gap \citep{Ruderman1975} or a partially screened inner gap \citep{Gil2003,Gil2006}.
The binding energy problem faced by the inner gap model (which was the motivation of the SCLF model) is alleviated if pulsars have strong surface magnetic field and low temperature \citep{Medin2007,Gil2006}, and can be completely solved if pulsars are bare strange quark stars \citep{Xu1999}. Therefore there is a strong motivation to study the oscillation effect on magnetospheric activities within the framework of the inner gap model, which is the subject of this paper.

Here we consider toroidal stellar oscillations. Similar to rotation, toroidal oscillations can also produce a voltage across pulsar polar cap due to unipolar induction, which would modify the Goldreich-Julian (1969) charge density and affect gap properties and particle acceleration processes. The sign of voltage produced by oscillations varies periodically. One may roughly estimate the period of an oscillation as the size of star over the speed of sound. The time scale of particle acceleration in the inner gap, on the other hand, is approximately the height of gap over speed of light. Since the size of star is much larger than the gap height and the speed of sound is much less than the speed of light, the time scale for particle acceleration is much shorter than the period of oscillation. Therefore, the periodic variation of the oscillation direction would not affect the ability of the inner gap to accelerate particles. Thanks to the additional voltage provided by oscillations, the properties of the inner gap, such as height and voltage, would be modified. In this paper, we present a general solution to determine the properties of the inner gap with oscillations. We consider three modes of pair plasma production: curvature radiation \citep[CR,][]{Ruderman1975}, inverse Compton scattering \citep[ICS,][]{Zhang1996}, and two-photon annihilation (2$\gamma$) processes \citep{Zhang1998}. We also discuss mode changing due to variation of the oscillation amplitude.

Interesting observational properties are observed to be associated with glitches in AXPs/SGRs.
AXPs/SGRs are isolated neutron stars with long periods ($2-10\,{\rm s}$) and high period derivatives ($10^{-13}-10^{-11}$). They are radio quiet in the quiescent state, which was considered as a distinctive characteristic with respective to normal radio pulsars \citep{Mereghetti2008}.
Recent observations suggest that some of them can become radio loud after some transient activities. Up to now, four AXPs/SGRs have been detected in the radio band: 1E 1547.0-5408, PSR J1622-4950, XTE J1810-197 and SGR J1745-2900 \citep{Olausen2014}\footnote{See McGill Online Magnetar Catalog:\\ http://www.physics.mcgill.ca/~pulsar/magnetar/main.html}, all of which behave as transient AXPs/SGRs. Pulsed radio emission from these objects was detected after some outbursts, which decays in a time scale of months to years\footnote{Transient, unpulsed, radio emission was also observed after the giant flares of SGR 1900+14 \citep{Frail1999} and SGR 1806-20 \citep{Gaensler2005,Cameron2005}, but those are believed to be the ``afterglow'' of the flaring material interacting with the ambient medium \citep[e.g.][]{Wang2005}.}. The origin of this transient radio emission of AXPs/SGRs is still a mystery, and some ideas to interpret it have been proposed \citep[e.g.][]{Morozova2012}. Since the outbursts of AXPs/SGRs sometimes are associated with glitches \citep{Kaspi2003,Dib2014}, it is possible that toroidal oscillations are excited during the outbursts, which modify the magnetospheric structure to allow radio emission to be produced. A direct motivation of our study is to explore this possibility.

In Section 2, we present the inner gap model under stellar oscillations. The general theory of the inner gap for three different modes are presented. In Section 3, we perform calculations of the gap properties by adopting typical parameters of radio pulsars and AXPs/SGRs. In Section 4, we apply the model to pulsars and AXPs/SGRs, and in particular, interpret the transient radio emission phenomenology of AXPs/SGRs.
Conclusions are drawn in Section 5 with some discussion.

\section{THE MODEL}

\subsection{Goldreich-Julian charge density and toroidal oscillation modification}
    Pulsars are rapidly rotating, highly magnetized compact objects.
Strong electric fields are generated
due to unipolar induction. \citet{Goldreich1969} found that in order to have the pulsar magnetosphere co-rotates with the star, the spatial electric charges have to follow a certain distribution, such that at every point in the magnetosphere the ${\bf E \times B}$ drift of the charges just maintain corotation of the magnetosphere. This charge density is called $\rho_{\rm GJ}$.
    Without the effect of oscillation and considering the general relativisty effect,  for a small-angle approximation $\theta \ll 1$, this density can be written as \citep{Muslimov1992,Morozova2010}
    \begin{align} \label{rho_GJ_rot}
        \rho_{\rm GJ,rot} &= -\frac{\Omega B}{2\pi c} \frac{1}{N\tilde{r}^3} \frac{f(\tilde{r})}{f(1)} \nonumber\\
        & \left[(1-\frac{\kappa}{\tilde{r}^3})\cos\alpha + \frac{3}{2}H(\tilde{r})\theta\sin\alpha\cos\phi \right].
    \end{align}
Here $\Omega$ is the rotation angular velocity of the pulsar, $B$ is the magnetic field strength at the pulsar surface, $c$ is the velocity of light, $\tilde{r}\equiv\frac{r}{R}$ is the reduced radial coordinate, $R$ is the radius of the pulsar, $\theta,\ \phi$ denote the angular coordinates in a spherical coordinate system with $z$-axis corresponding to the spin axis, $\alpha$ is the magnetic axis inclination angle of the pulsar, $N=\sqrt{1-\frac{2GM}{c^2r}}$ is the lapse function, $\kappa\equiv\frac{2GM}{Rc^2}\frac{I}{MR^2}$, and $M$ and $I$ are the mass and moment of inertia of pulsar, respectively.
    The function $f(\tilde{r})$ reads
    \begin{equation}\label{f}
        f(\tilde{r})=-3\left(\frac{\tilde{r}}{\varepsilon}\right)^3 \left[ \ln(1-\frac{\varepsilon}{\tilde{r}}) + \frac{\varepsilon}{\tilde{r}}(1+\frac{\varepsilon}{2\tilde{r}}) \right],
    \end{equation}
    where $\varepsilon=2GM/Rc^2$, and the function $H(\tilde{r})$ reads
    \begin{equation}
        H(\tilde{r}) = \frac{1}{\tilde{r}}\left(\varepsilon-\frac{\kappa}{\tilde{r}^2}\right) + \left(1-\frac{3}{2}\frac{\varepsilon}{\tilde{r}}+\frac{1}{2}\frac{\kappa}{\tilde{r}^3}\right) \left[f(\tilde{r})(1-\frac{\varepsilon}{\tilde{r}})\right]^{-1}.
    \end{equation}

    We now consider the effect of oscillation. In general, there are two types of oscillations: spherical and toroidal. Only the toroidal oscillations can effectively modify the spin of the star, and therefore affect the Goldrich-Julian charge density and magnetospheric configurations. Spherical oscillations, on the other hand, would not modify the magnetosphere structure significantly.
Therefore here we only consider toroidal oscillations. The veolocity components of toroidal oscillations can be written as \citep{Unno1989}
    \begin{equation} \label{oscillation}
        \delta v^{\hat{i}} = \left[0,\  \frac{1}{\sin\theta}\partial_{\phi} Y_{lm}(\theta,\phi),\  -\partial_\theta Y_{lm}(\theta,\phi) \right] \tilde{\eta}(\tilde{r})e^{-i\omega t},
    \end{equation}
    where
    $Y_{lm}$ is the spherical harmonic function, $\tilde{\eta}(\tilde{r})$ is a parameter denoting the amplitude of the oscillations, and $\omega$ is the angular frequency of the oscillations. Generally speaking, the axis of toroidal oscillations should be close to the spin axis itself, so the $z$-axis of the spherical coordinate system is likely the spin axis.
Because the height of a pulsar inner gap is much smaller than the pulsar radius, in our calculations we take $\tilde{\eta}(\tilde{r})=\tilde{\eta}(1)$.

    \citet{Morozova2010} obtained the modifed G-J charge density caused by oscillations (their Eq. (13)). However, there is a mistake in their derivation from their Eq. (11) to Eq. (13). The factor $1/\Theta^2(\tilde{r})$ ($\Theta$ is the co-latitude of the last closed magnetic field line) should not appear in the second term of their Eq. (13). Within the polar cap region, where the co-latitude with respective to the magnetic axis $\vartheta$ is small, the correct expression should read
    \begin{equation} \label{rho_GJ_osc}
        \rho_{\rm GJ,osc} = -K \frac{\Omega B}{4\pi c} \frac{1}{\tilde{r}^4} \frac{1}{N} \frac{f(\tilde{r})}{f(1)} l(l+1) Y_{lm}(\theta,\phi) \cos\vartheta e^{-i\omega t},
    \end{equation}
    where $K=\tilde{\eta}(1)/\Omega R$ denotes the ratio of oscillation velocity and rotation velocity in the pulsar surface, which represents the intensity of the oscillations.
    Since
$\Theta$ is a small quantity at the polar cap region, the factor $1/\Theta^2(\tilde{r})$ is a large quantity. As a result, the $\rho_{\rm GJ,osc}$ value obtained in their paper is significantly over-calculated. After correcting the error, one gets the ratio
    \begin{equation} \label{rho_GJ_ratio}
        \frac{\rho_{\rm GJ,osc}}{\rho_{\rm GJ,rot}} = \frac{K}{2\tilde{r}} \frac{l(l+1)Y_{lm}(\theta,\phi)\cos\vartheta}{1-\kappa/\tilde{r}^3} e^{-i\omega t}.
    \end{equation}

With $z$ axis fixed to the spin axis, the pulsar polar cap region would have $\theta \sim \alpha$, where $\alpha$ is the magnetic axis inclination angle of the pulsar. For a non-extreme $\alpha$ in the realistic cases, given a same value of $K$, the oscillation modes with a same $l$ but different $m$ values would have $\rho_{\rm GJ}$ values of the same order of magnitude. For different $l$ values, $\rho_{\rm GJ}$ increases by a factor $Kl(l+1)Y_{lm}$, and the oscillation velocity increase as a factor $\sim KlY_{lm}$. As a result, when $l$ is large, a large $\rho_{\rm GJ}$ would be obtained with a relatively smaller oscillation velocity.

It is unclear in reality which mode(s) would be excited and be dominant. Practically, introducing the inclination angle $\alpha$ would significantly complicate the calculations, so in the following calculations we assume $\alpha=0$ for simplification, so that $\theta=\vartheta$. At this configuration, the $m=0$ modes have similar amplitudes as the cases with an arbitrary $\alpha$ except those close to 90 degree, and we only calculate $l=2,\ m=0$ mode as an illustration\footnote{Since $Y_{lm}(\theta,\phi)$ has a factor of $\sin^m\theta$, the $m>1$ modes are much suppressed when $\theta \sim 0$, so that $\alpha=0$ is a bad approximation.}. As discussed above, this example may be regarded as a typical example of $l=2$ for arbitrary $\alpha$ values except those close to 90 degree and arbitrary $m$ values, and the cases for different $l$ values may be inferred through the simple scaling relations.

    \subsection{Basic physical picture of inner gap}
The picture of the vacuum gap model (RS75) is the following. Due to a centrifugal force, particles in a pulsar magnetosphere continuously streams out the light cylinder in the open field line regions. At the polar cap of an ``anti-parallel'' rotator, if the ion binding energy at the pulsar surface is large enough, no charge can be provided to sustain a continuous outflow, and a gap, in which the charge density $\rho \ll \rho_{\rm GJ}$, is formed. An electric field component parallel to the magnetic field, $E_\parallel$, is developed.
Seed $\gamma$-rays in the gap would materialize in the strong $B$ field and generate electron-positron pairs. These pairs are accelerated in the gap to ultra-relativistic energies and produce high-energy photons through CR or ICS. High-energy photons undergo pair production through $\gamma-B$ or two-photon processes, and the newly produced positrons/electrons are accelerated and radiate again. This leads to a run-away pair creation avalanch, which ultimately leads to discharge of the gap potential and breakdown of the gap.  Denoting the gap height as $h$, under one-dimensional approximation for $h\ll r_p$ (where $r_p=R(\frac{\Omega R}{c})^{1/2}$ is the polar cap radius), one can write the electric field in the gap
    \begin{equation}
        E(s)=-4\pi\int_h^s\rho_{\rm GJ}(s){\rm d}s,
    \end{equation}
    where $s$ denotes the distance from the surface of the star. The corresponding electric potential difference  along the gap is
    \begin{equation}
        \Delta V(h) = -\int_0^h E(s) {\rm d}s.
    \end{equation}

    There is a maximum voltage across the polar cap provided by the unipolar induction, denoted as $\Delta V_{\rm max}$. When $\Delta V$ approaches $\Delta V_{\rm max}$, the 1D approximation above would fail.
The criterion
    \begin{equation} \label{death line}
        \Delta V(h) < \Delta V_{\rm max}
    \end{equation}
ensures that pair production screens $E_\parallel$ and limits the gap potential. Since pair production is the necessary condition for coherent radio emission,
Eq. (\ref{death line}) defines the radio death line of pulsars \citep[e.g. RS75,][]{Zhang2000}.

    For $\alpha=0$, the maximum voltage across the polar cap can be written as (see derivation in the Appendix)
    \begin{align}\label{V_max_m=0}
        \Delta V_{\rm max} & \approx -\frac{\Omega R^2NB_0}{c} \frac{\Theta_0^2}{2} \nonumber\\
        & \left[(1-\kappa) + K\frac{l(l+1)}{2}Y_{lm}(\Theta_0,\phi)\right],~{\rm for}~ m=0;
    \end{align}
    \begin{align}\label{V_max_m!=0}
        \Delta V_{\rm max} & \approx -\frac{\Omega R^2NB_0}{c} \nonumber\\
        & \left[(1-\kappa)\frac{\sin^2\Theta_0}{2} - KY_{lm}(\Theta_0,\phi)\right],
        {\rm for}\ m\neq0.
    \end{align}
    And the polar angle of the last open magnetic field line $\Theta_0$ is determined by
    \begin{equation} \label{polar_angle_m=0}
        \Theta_0 \approx \frac{\sqrt[4]{N}}{f(1)} \sqrt{\frac{2\Omega R}{c} \left[(1-\kappa)+K\frac{l(l+1)}{2}Y_{l0}\right]},\ {\rm for}\ m=0;
    \end{equation}
    \begin{align} \label{polar_angle_m!=0}
        \frac{1}{2}f^4(1)\Theta_0^6 \approx \frac{\Omega^2R^2N}{c^2} \left[(1-\kappa)\frac{\sin^2\Theta_0}{2}-KY_{lm}(\Theta_0,\phi)\right] \nonumber\\ \left[(1-\kappa)+K\frac{l(l+1)}{2}Y_{lm}(\Theta_0,\phi)\right],\ {\rm for}\ m\neq0.
    \end{align}

    \subsection{Different gap modes}

    For a pulsar inner gap, several processes operate simultaneously. Gamma-rays are produced via both CR and ICS by electrons/positrons, and pair productions proceed through both $\gamma-B$ and $2\gamma$ processes. The height and potential of the gap is determined by the process which takes the shortest distance to generate enough pairs to shut down the gap \citep{Zhang1997}. Based on physical conditions, one of the following three modes may play a dominant role to define gap parameters: CR, resonant ICS, and $2\gamma$ processes. In the following, we discuss these modes in turn.

\subsubsection{CR mode}

    In RS75, only the CR mode was considered, and the height of gap is determined by $h\sim l_\gamma$, where $l_\gamma$ is the mean free path of typical $\gamma$ photons, which reads
    \begin{eqnarray}\label{l_gamma}
        l_\gamma = \frac{4.4}{e^2/\hbar c} \frac{\hbar}{m_e c} \frac{B_q}{B_\bot}\exp(\frac{4}{3\chi}), \
        \chi \ll 1, \\
        \chi \equiv \frac{E_\gamma}{2m_e c^2} \frac{B_\bot}{B_q} \equiv \frac{E_{\gamma\bot}}{2m_e c^2} \frac{B}{B_q}.
    \end{eqnarray}
    where $e$ and $m_e$ are the electric charge and the mass of an electron, respectively, $c$ is speed of light, $\hbar$ is reduced Planck's constant, $B_q=m^2 c^3/e\hbar=4.4\times10^{33}\,{\rm G}$ is the critical magnetic field, $B_\bot$ is the perpendicular magnetic component with respect to the direction of the $\gamma$-ray photon in the co-rotating frame, $E_\gamma$ is the energy of the $\gamma$-ray photon, and $E_{\gamma\bot}$ is the perpendicular energy of the $\gamma$ photon with respect to the direction of the magnetic field in the co-rotation frame. According to RS75, a photon emitted parallel to the direction of the local magnetic field line and propagates along a distance $h$ would see a perpendicular magnetic field
    \begin{equation}
        B_\bot=\frac{h}{\rho}B,
    \end{equation}
    where $\rho$ is the curvature radius of the magnetic field line.
    For a dipolar magnetic field, one has
    \begin{equation}
        \rho \simeq \frac{4}{3} \sqrt{Rc/\Omega} \simeq 9.21\times10^7(P R_6)^{1/2}\,{\rm cm},
    \end{equation}
    where $P$ is the rotational period of pulsar, $R_6$ is pulsar's radius in units of $10^6\,{\rm cm}$.
    For multipole magnetic fields, one may adopt a typical curvature radius $\rho\sim10^6\,{\rm cm}$, comparable to the size of the pulsar. We consider both possibilities in the following calculations.

    In order to break a gap, a lepton needs to at least release one photon within a typical mean free path $l_e$, and the photon needs to attenuate in the magnetic field to produce pairs within a mean free path $l_\gamma$ (Eq.(\ref{l_gamma})). For the CR process, $l_e \ll l_\gamma$ is usually satisfied \citep{Zhang2000}, one may then take $h \sim l_\gamma$. Notice that in Eq. (\ref{l_gamma}), $l_\gamma$ decreases with increasing $E_\gamma$, so the largest Lorentz factor of electrons/positrons $\gamma_e$ determines the height of gap. Under the assumption that electrons/positrons can be accelerated to the largest Lorentz factor in the gap $\gamma_e m_e c^2=\Delta V(h)$, one can solve the height of gap with Eq. (\ref{l_gamma}) (RS75), denoted as $h_{\rm min}$.

    However, in some situations, especially when the ICS mode is invoked,
$l_e$ often exceeds $h_{min}$ (see more detail below). As a result, $l_\gamma$ related to the largest $\gamma_e$ is no longer a good estimate of the gap height $h$, since at this distance, an electron on average has not produced a photon yet. The photon-pair cascade chain is not launched. In this case,
we determine the gap height via
    \begin{equation}\label{h_equal}
        l_e(\gamma_e) = l_\gamma(\gamma_e) = h(\gamma_e).
    \end{equation}
    When oscillations are considered, in some cases, even the CR mode also has $l_e > l_\gamma$.

    For the CR mode, we have
    \begin{equation}
        E_{\gamma,{\rm CR}}=\hbar\omega_{c,{\rm CR}}=\hbar\frac{3}{2}\frac{\gamma_e^3 c}{\rho},
    \end{equation}
    \begin{equation}
        l_{e,{\rm CR}} \sim c \left( \frac{P_{\rm CR}}{\hbar\omega_{c,{\rm CR}}} \right)^{-1}
        = \frac{9}{4}\frac{\hbar\rho c}{\gamma_e e^2},
    \end{equation}
    where $\gamma_e$ is the Lorentz factor of electrons/positrons.

\subsubsection{ICS mode}\label{sec:ICS}

    For the ICS mode, we mainly focus on ICS at resonance. The scattering cross section of ICS becomes very large when the incident photon energy in the electron rest frame equals to the energy difference between electron's Laudau level interval from the ground state to the first excited state \citep{Xia1985}.
The large scattering cross section of this resonant mode makes $l_e$ reduce significantly so that pair cascade in the gap becomes possible \citep{Zhang1996}. There is another typical ICS energy ($\gamma_e^2 kT$), which is related to boosting the thermal peak of the source photons to high energies. This mode, called the thermal mode \citep{Zhang1997}, was later found to have too large a $l_e$, which may not play an important role \citep{Zhang2000}. We do not consider this mode in the following calculations.

    In the resonant ICS mode, when the magnetic field is below critical magnetic field, we have \citep{Zhang1997}
    \begin{equation} \label{E_gamma_res}
        E_{\gamma,{\rm res}} \sim 2\gamma_e^2\hbar\omega_{\rm res}(1-\beta_e\mu_i) = 2\hbar\gamma_e\omega_B,
    \end{equation}
    where $\beta_e=(1-\gamma_e^{-2})^{1/2}$ is the dimensionless velocity of the relativistic particle, $\mu_i$ is the cosine of the angle between incident photon and the direction of electron's velocity, $\omega_B=eB/mc$ is the cyclotron frequency of an electron/positron in the magnetic field, and $\omega_{\rm res}=\omega_B/[\gamma_e(1-\beta_e\mu_i)]$ is the so-called resonant frequency of the ICS process.
    When the magnetic field exceeds $B_q$, the resonant ICS is in the Klein-Nishina regime, so that the largest energy of scattered photon is essentially the total energy of electron
    \begin{equation}
        E_{\gamma,{\rm max}} \simeq \gamma_e m_e c^2.
    \end{equation}
    For given energy of scattered photon $\epsilon_s$ (in units of $m_e c^2$), the incident photon's energy (in units of $m_e c^2$) is
    \begin{equation}
        \epsilon_i=\epsilon_s/[2\gamma_e^2(1-\beta_e\mu_i)].
    \end{equation}
    Then the mean free path of an electron can be written as
    \begin{align}
        l_{e,{\rm ICS}}(\epsilon_s) &\sim
        [ \int_0^1 \sigma^\prime(\epsilon_s, \mu_i) (1-\beta_e\mu_i) n_{\rm ph}(\epsilon_s, \mu_i) \nonumber\\
        & \times\frac{\epsilon_s}{2\gamma_e^2} \frac{\beta_e}{(1-\beta_e\mu_i)^2} d\mu_i ]^{-1}
    \end{align}
    where
    \begin{equation}
        n_{\rm ph}(\epsilon_i)d\epsilon_i = \frac{8\pi}{\lambda_c^3} \frac{\epsilon_i^2}{\exp(\epsilon_i/\theta_{\rm T})-1} d\epsilon_i
    \end{equation}
    represents the photon number density distribution of a semi-isotropic blackbody radiation, in which  $\theta_{\rm T}=k_B T/m_e c^2$, $k_B$ is Boltzmann's constant, and $\lambda_c=\hbar/m_e c \simeq 2.426\times10^{-10}\,{\rm cm}$ is the electron Compton wavelength.
    Here $\sigma^\prime$ is the scattering cross section in the electron rest frame (ERF), which is calculated using the Sokolov \& Ternov cross section in \citet{Baring2011}
    \begin{align}
        \frac{d\sigma^\prime_{\rm ST}}{d(\cos\theta_f)} &=  \frac{3\sigma_T}{64} \frac{\omega_f^2e^{-\kappa}}{\omega_i[2\omega_i-\omega_f-\zeta]\varepsilon_\perp^3} \nonumber\\
        &\times \sum_{s=\pm1}\frac{(\varepsilon_\perp-s)^2 \Lambda_s}{(\omega_i-B/B_q)^2+(\Gamma_s/2)^2}.
    \end{align}
    Here $\theta_f$ is the scattered angle in the ERF, $\sigma_T$ is the Thomson cross section, $\omega_i=\gamma_e\epsilon_i(1-\beta_e\mu_i)$ is the incident photon energy in units of $m_e c^2$ in the ERF, $\omega_f$ is the energy of scattered photon in the ERF written as
    \begin{equation}
        \omega_f = \frac{2\omega_i r}{1+\sqrt{1-2\omega_i r^2 \sin^2\theta_f}}, \ r=\frac{1}{1+\omega_i(1-\cos\theta_f)},
    \end{equation}
    $\varepsilon_\perp=\sqrt{1+2B/B_q}$, $\kappa=\omega_f^2\sin^2\theta_f/(2B/B_q)$, $\zeta=\omega_i\omega_f(1-\cos\theta_f)$, and
    \begin{equation}
        \Lambda_s = (2\varepsilon_\perp+s){\omega_f^2\mathcal{T}-[\omega_i-\omega_f]} - s\varepsilon_\perp^2[\omega_i-\omega_f],
    \end{equation}
    \begin{equation}
        \omega_f^2\mathcal{T} = 2(\omega_i-\omega_f)-2\omega_i\omega_f(1+\omega_i){1-\cos\theta_f}+2\omega_i^2,
    \end{equation}
    \begin{equation}
        \Gamma_s = \left(1-\frac{s}{\varepsilon_\perp}\right)\Gamma, \ \Gamma=\alpha_f\frac{B}{B_q}I_1(B),
    \end{equation}
    \begin{align}
        I_1(B) &= \int_0^\Phi \frac{d\phi e^{-\phi}}{\sqrt{(\Phi-\phi)(1/\Phi-\phi)}} \left[1-\frac{\phi}{2}(\Phi+1/\Phi)\right], \nonumber\\
        \Phi &= \frac{\sqrt{1+2B/B_q}-1}{\sqrt{1+2B/B_q}+1}.
    \end{align}
    The reason why the mean free path of electron would exceed that of photons in resonant ICS mode is ascribed to the offset between the energy of the resonant incident photons $\epsilon({\rm res})$ (which provide the largest cross section) and the energy of the peak of blackbody radiation (which provide the largest number of photons).

    When magnetic field exceeds the critical magnetic field $B_q$, the resonant condition cannot be satisfied since the resonant photons are in the Klein-Nishina regime. The ICS cross section declines, so that the electron mean free path $l_e$ becomes much larger. In such a case, no solution can be found for Eq. (\ref{h_equal}) for resonant ICS mode. In any case, one may still have an ICS mode, even though the photon energies are not at the resonance. We then apply the minimum $l_e(\gamma_e)$ to define the gap height.

    For the resonant ICS mode, considering that the energy loss of particles in the acceleration process cannot exceed the maximum electric potential energy in the gap, we need to ensure that the voltage  along the gap is enough to accelerate particles to the Lorentz factor required. This condition can be expressed as \citep{Zhang1996}
    \begin{equation}\label{P_e}
        P_e \frac{h}{c}<e \Delta V(h).
    \end{equation}
    We adopt an approximation of the energy-loss rate at resonance \citep{Dermer1990}:
    \begin{equation}
        P_e=\frac{c\sigma_{\rm eff}}{\mu_+-\mu_-}(\frac{\epsilon_B^2}{\gamma_e}) \times\int_{\epsilon_B/\gamma_e(1-\beta_e\mu_-)}^{\epsilon_B/\gamma_e(1-\beta_e\mu_-)} d\epsilon \epsilon^{-2} n_{\rm ph}(\epsilon)(1-\frac{\epsilon}{\gamma_e\epsilon_B}),
    \end{equation}
    where $\sigma_{\rm eff}\approx 3\pi\sigma_T\alpha_f$ is the effective cross section, $\mu_+$, $\mu_-$ are cosines of the maximum and minimum incident angles, respectively, which we adopt as $\mu_\pm=\pm 1$ in the calculations.
    In most cases, such a requirement is satisfied.

\subsubsection{Two photon pair production mode}

    As for the $2\gamma$ process, we consider interaction between the photons created by high-energy particles (through CR or ICS) and thermal photons. The mean free path of a photon, $l_{2\gamma}$, is determined by \citep{Zhang1998}
    \begin{equation}
        \int_0^{l_{2\gamma}} 2.1\times10^{-6}T_6^3 {\rm Max}[f(\nu, x)]dx=1,
    \end{equation}
    \begin{equation}
        {\rm Max}[f(\nu, x)]=0.270-0.507\mu_c+0.237\mu_c^2,
    \end{equation}
where $T_6$ is the surface temperature of pulsar in units of $10^6\,{\rm K}$, $\mu_c$ is the cosine of the angle between the two photons. Since gap height is usually much shorter than the size of polar caps, most of the two-photon interactions occur at about $90^{\rm o}$ angle (one photon streaming out, while the other comes perpendicularly from other parts of the surface), so we simply adopt $\mu_c \sim 0$. A small change of $\mu_c$ would not noticeably affect the results. It is worth pointing out that the height of gap due to $2\gamma$ processes is only relative to the surface temperature and has nothing to do with the expression of $V(h)$; in other words, oscillations do not have an essential effect on $2\gamma$ processes. The $2\gamma$ process is more important in SGRs/AXPs thanks to their high surface temperatures \citep{Zhang2001}.

\section{EFFECT OF OSCILLATIONS ON INNER GAP PROPERTIES}
When an oscillation gets its maximum amplitude with the same direction of rotation, it has the maximum effect on the properties of the gap. If pair production condition is satisfied at this most optimistic condition, radio emission can be produced. Therefore in the following we mainly focus on this extreme condition and ignore the periodical variation of the oscillations. In other words, in Eq. (\ref{rho_GJ_osc}) we set $e^{-i\omega t}=1$. In principle the gap properties would vary with time during one oscillation, but it is difficult to have an observational effect due to the high oscillation frequency. We may estimate the frequency of oscillation in a pulsar based on its analogy with the earth. The pulsar oscilation frequency is $P_{\rm osc} \propto \frac{L}{v}$, where $L$ is the typical length of the star which is or the order of pulsar radius $\sim 10^6\,{\rm cm}$, $v$ is the speed of sound, which might be close to speed of light for both neutron stars and solid quark stars \citep[e.g.][]{Haensel2007}. From the observations of earth oscillations, we know that the period of its fundamental-frequency oscillation is $P_{\rm earth}\sim 1\,{\rm h}$. The radius of the earth is $R_{\rm earth}\sim 6.4\times10^8\,{\rm cm}$, and the speed of sound (e.g. that earthquake wave) is $v_{\rm earth} \sim 5\times10^5\,{\rm cm/s}$.
As a result, the period of pulsar oscillation may be estimated as
\begin{equation}
    P_{\rm osc} \sim \frac{R}{R_{\rm earth}} \frac{v_{\rm earth}}{v} P_{\rm earth} \sim 3\times10^{-4}\,{\rm s}.
\end{equation}
The frequency is
\begin{equation}
    \nu_{\rm osc} = \frac{1}{P_{\rm osc}} \sim 3\,{\rm kHz}.
\end{equation}
This is much higher than the rotational frequency, so that the periodical variation of $\rho_{\rm GJ,osc}$ would be smoothed out and cannot cause an observable, periodical signal.
In principle an integrated average effect may be arguably more relevant, but this involves much more complicated calculations. The treatment here can provide an estimate to correct order of magnitude. Moreover, since we will discuss the threshold condition for pair production (and hence, radio emission), the maximum potential would be more relevant to define such a condition.

\begin{figure*}
    \centering
    \includegraphics [width=15cm]{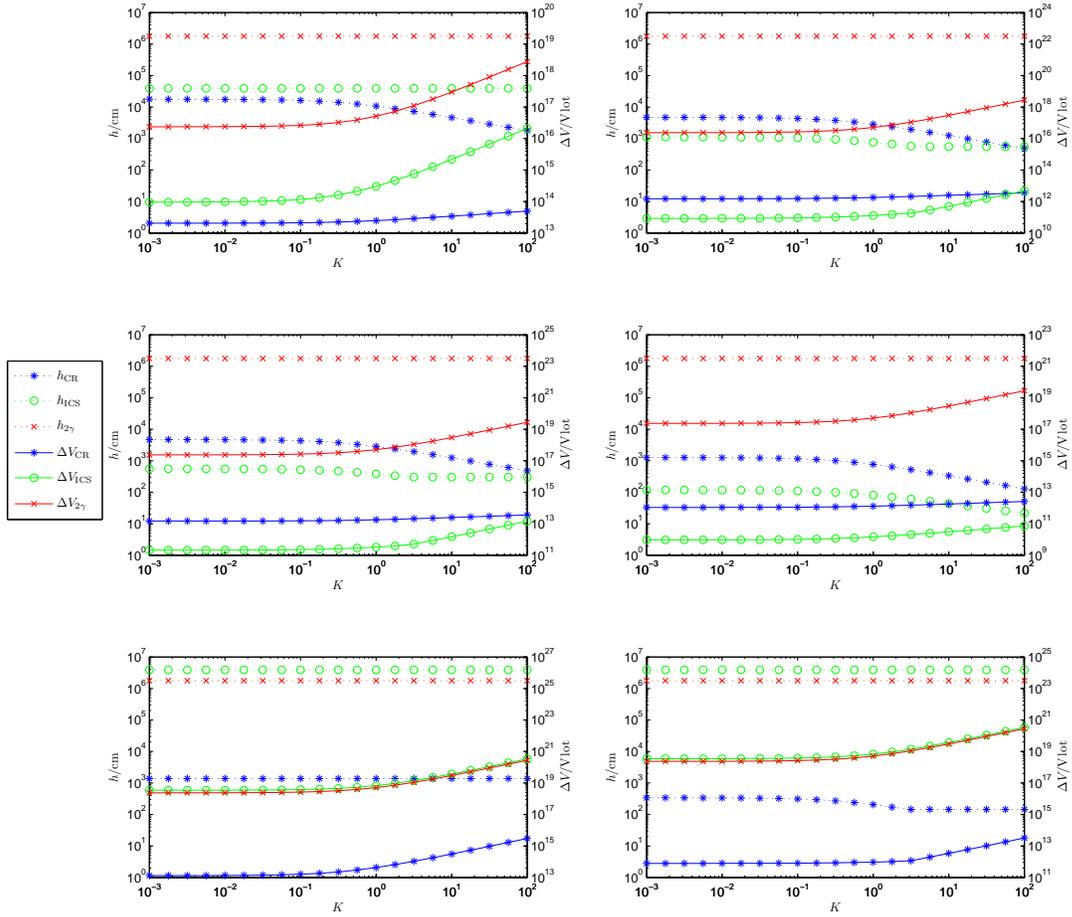}
    \caption{\label{P=1_T6=1_l=2_m=0} Gap properties as a function of $K$, the dimensionless amplitude of oscillations, which approximately equals to the ratio of oscillation velocity and rotational velocity in the surface of the pulsar. The dotted lines represent the gap heights, while the solid lines represent the voltages along the gap. The blue lines with star points, green lines with circle points, and red lines with cross points represent modes of curvature radiation (CR), inverse Compton scattering (ICS) and two-photon progress ($2\gamma$), respectively. Here $P=1\,{\rm s}$, $T=10^6\,{\rm K}$ and the $l=2,\ m=0$ oscillation mode are adopted. The top two. middle two, and bottom two panels are for $B=10^{12}\,{\rm G}$, $10^{13}\,{\rm G}$, $10^{14}\,{\rm G}$, respectively. The left three are for a dipole magnetic field configuration, whereas the right three are for a multipole configuration.}
\end{figure*}

    We show the general effects of oscillation for a typical pulsar period $P=1\,{\rm s}$ in Fig. \ref{P=1_T6=1_l=2_m=0}.
The temperature $T=10^6\,{\rm K}$, and magnetic field strengths $B=10^{12},10^{13},10^{14}\,{\rm G}$, are adopted, respectively. Both dipole and multipole magnetic configurations are considered. The point lines represent the gap heights, while the solid lines represent the voltages along the gap. The blue lines with star points, green lines with circle points and red lines with cross points represent CR, ICS, and $2\gamma$ modes, respectively. The gap properties are plotted against the dimensionless amplitude of oscillation $K$, which approximately equals to the ratio of oscillation velocity to rotational velocity at the surface of star. Our calculations start from $K=10^{-3}$, at which the differences of gap properties are small enough with respect to the cases without oscillation.  Note that all the cases we calculate satisfy the requirement in Eq. (\ref{P_e}), so we do not present the condition in the figures.

\begin{figure*}
    \centering
    \includegraphics [width=15cm]{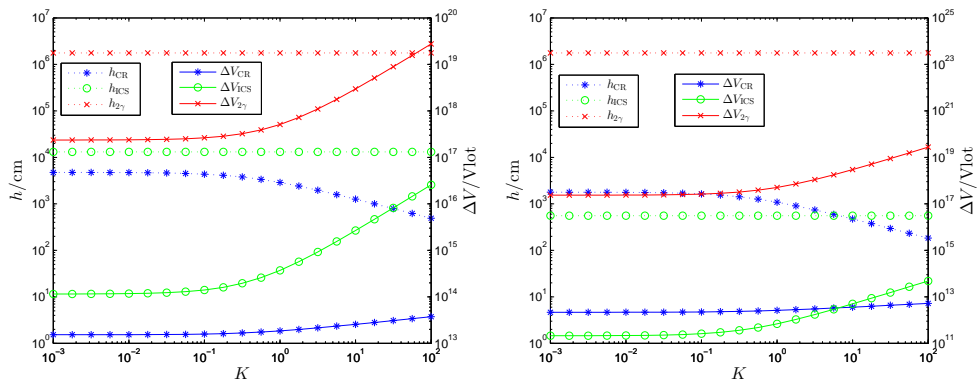}
    \caption{\label{P=0.1_B12=1_T6=1_l=2_m=0} Gap properties as a function of dimensionless oscillation amplitude $K$ for a typical radio pulsar with $P=0.1$ s, $T=10^6$ K, and $B=10^{12}$ G. The left and right panels are for the dipole and multipole configurations, respectively. Same line conventions as Fig.\ref{P=1_T6=1_l=2_m=0}.}
\end{figure*}

\begin{figure*}
    \centering
    \includegraphics [width=15cm]{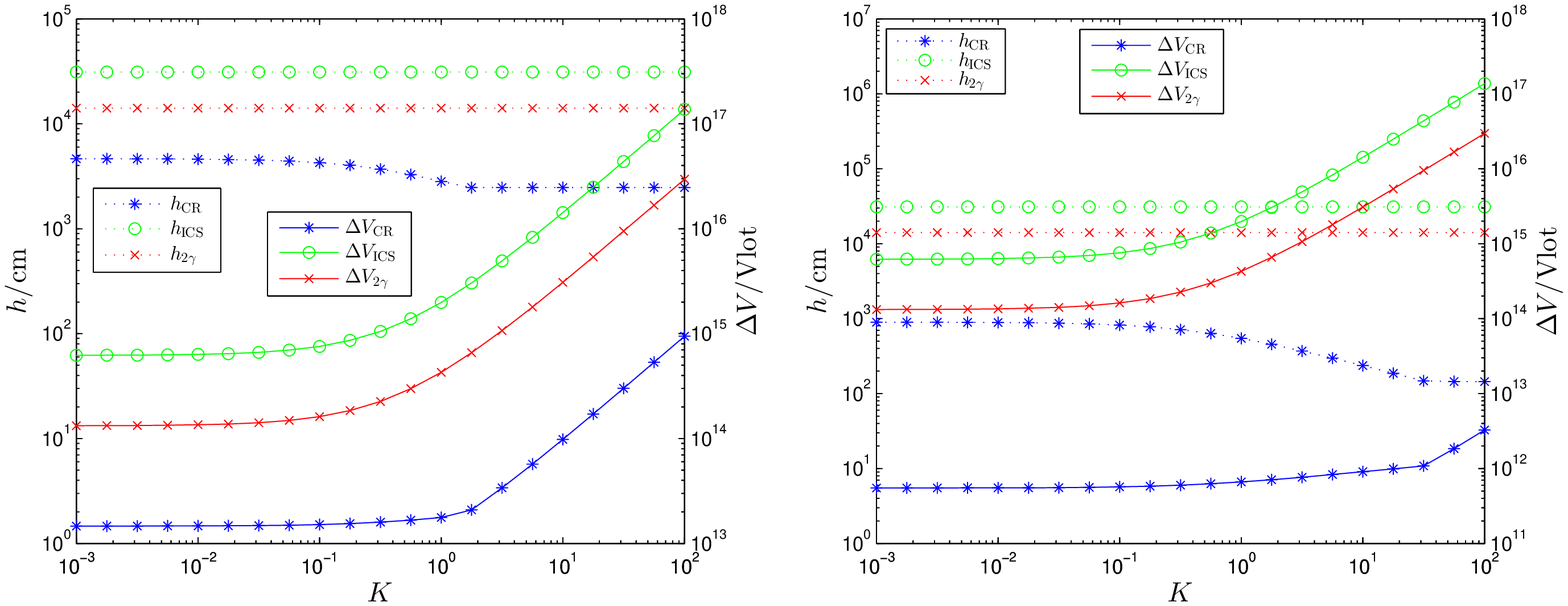}
    \caption{\label{P=10_B12=100_T6=5_l=2_m=0} Gap properties as a function of dimensionless oscillation amplitude $K$ for a typical magnetar with $P=10$ s, $T=5\times 10^6$ K, and $B=10^{14}$ G. The left and right panels are for the dipole and multipole configurations, respectively. Same line conventions as Fig.\ref{P=1_T6=1_l=2_m=0}.}
\end{figure*}

   Keeping other parameters the same, we calculate a faster pulsar with $P=0.1$ s in Fig. \ref{P=0.1_B12=1_T6=1_l=2_m=0}. A typical SGR/AXP with $P=10$ s, $T=5\times 10^6$ K, and $B = 10^{14}$ G is calculated in Fig. \ref{P=10_B12=100_T6=5_l=2_m=0}.

    As mentioned above, oscillations have the effect to enhance the voltage across the polar cap when we have the same gap height, which means that $\Delta V(h)$ increases. For the $2\gamma$ mode, the height of gap is only determined by temperature, therefore it remains unchanged as the oscillation amplitude and $\Delta V$ vary.
The CR and resonant ICS modes, on the other hand, are senstive to these variations. The gap height may decrease when oscillations occur. Notice that a larger voltage across the polar cap would accelerate particles to higher Lorentz factors, and photons emitted by high energy particles would have a higher typical energy, hence, a shorter mean free path, so that the gap would break down earlier and has a lower height. In this case, $\Delta V$ still increases but with a smaller slope than the $2\gamma$ mode case.

    In some cases the gap height in the CR and resonant ICS modes would also remain unchanged. These cases occur when $l_e > h_{\rm min}$ ($h_{\rm min}$ is the mean free path of photons when $\gamma_e=e\Delta V/(m_e c^2)$), so that the height of gap should be determined by Eq. (\ref{h_equal}). Notice that $\Delta V(h)$ does not appear in Eq. (\ref{h_equal}) or the expression of $l_e$, which means that the gap height does not react to oscillations when $l_e > h_{\rm min}$, so that the gap height cannot be shorter than a lower limit determined by Eq. (\ref{h_equal}) or a minimum of $l_e$. In the resonant ICS mode, $l_e> h_{\rm min}$ occurs more frequently (as discussed in \S\ref{sec:ICS}).

    Notice that in reality the gap properties are determined by the mode whose gap height is the lowest.
 As a result, mode changing can happen as the oscillation amplitude varies, or other parameters (e.g. surface temperature or magnetic field strength) change (see figures below). \cite{Zhang1997} first discussed such gap mode changing due to fluctuation of the pulsar surface temperature.

    The $2\gamma$ mode dominates only in high temperature and a relative small $V(h)$,  because a high temperature makes $2\gamma$ process more effective and a relative small $V(h)$ makes the CR and ICS modes less effective. Here a relative small $V(h)$ means a weak magnetic field, a large period without oscillations.
    When $V(h)$ is large, the CR and ICS modes would become more effective, therefore the $2\gamma$ process could not dominate, especially with oscillations. Comparing the CR mode with the resonant ICS mode, $l_\gamma$ in the resonant ICS mode is much shorter than that in the CR mode. However, when the magnetic field strength exceeds $B_q$, or is too weak with a dipole configuration, $l_e > h_{\rm min}$ often occurs in the resonant ICS mode, so that $l_e$ defines the gap height, which is higher.
In these cases, the CR mode would dominate the properties of gap in these cases. The ICS mode plays an important role only when the magnetic field does not exceed $B_q$, typically with a multipole configuration, but sometimes in a strong enough dipole field also.

The influence of different magnetic field configurations is reflected in the curvature radii of magnetic field lines. The dipole magnetic field lines have larger curvature radii, which make the gap heights in the CR and resonant ICS modes higher. It also influences the condition of $l_e > h_{\rm min}$.
%\newpage

After general considerations of the vacuum gap, we also analyze the oscillations effect on partially screened inner gaps. Following \cite{Gil2003,Gil2006}, we introduce a shielding factor defined as $\eta \equiv 1-\rho_i/\rho_{\rm GJ}$ where $\rho_i$ is the real electric charge density in the gap contributed by iron ions. We determine the shielding factor $\eta$ and surface temperature $T_s$ by \citep{Gil2006}
\begin{equation}
    \sigma_{\rm SB}T_s^4 = \gamma_e m_e c^3 \frac{\rho_{\rm GJ}}{e} \eta,
\end{equation}
\begin{equation}
    \eta \approx 1- \exp(30-\frac{\varepsilon_c}{k_B T_s}) \frac{\rho_{\rm GJ,rot}}{\rho_{\rm GJ}}, \
    \varepsilon_c \approx (0.18\,{\rm keV})(\frac{B}{10^{12}\,{\rm G}})^{0.7}
\end{equation}
where $\sigma_{\rm SB}$ is the Stefan-Boltzman constant and $\varepsilon_c$ is the binding energy in the neutron star surface.

\begin{figure*}
    \centering
    \epsscale{0.8}
    \plotone{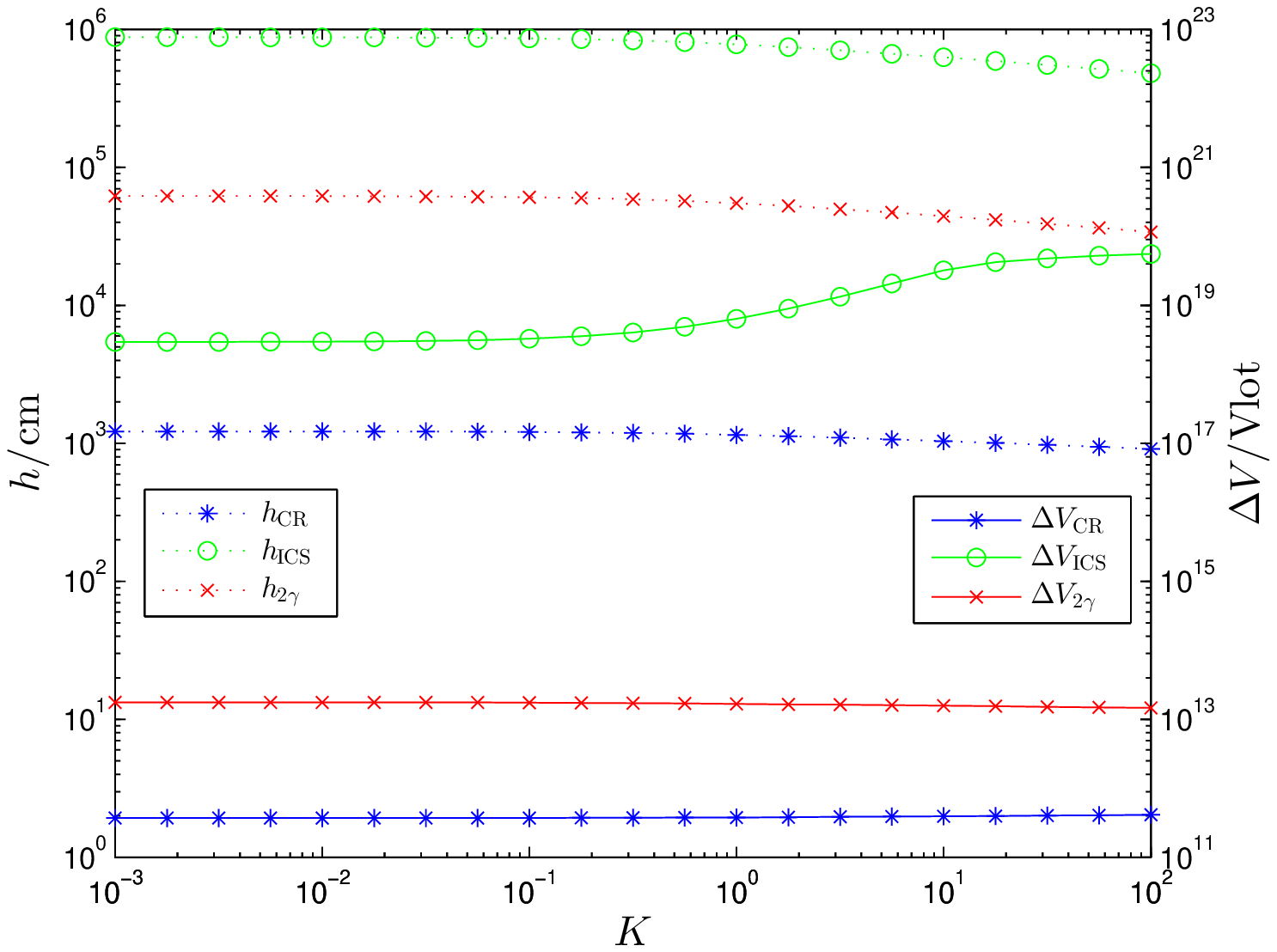}
    \caption{\label{PSG_5B_q} The properties of a partially screened gap as a function of the dimensionless oscillation amplitude $K$. $P=1\,{\rm s}$,$B_{\rm di}=10^{12}\,{\rm G}$ and $B_{\rm multi}=5B_q$ are adopted. Same line conventions as Fig.\ref{P=1_T6=1_l=2_m=0}.}
\end{figure*}

In Fig. \ref{PSG_5B_q}, we calculate the properties of the partially screened gap as a function of the dimensionless oscillation amplitude $K$. Here we adopt
$P=1\,{\rm s}$,
the dipole magnetic field $B_{\rm di}=10^{12}\,{\rm G}$ and the multipole magnetic field $B_{\rm multi}=5B_q$, as is expected in the partially screened gap scenario. One can see that although oscillations increase the GJ charge density and are able to increase the voltage along the gap, they also enhance the heating by the back flowing particles and thus decrease the shielding factor. As a result, the voltage increase is not as obvious in partially screened gaps as in vacuum gaps.

\section{APPLICATIONS}

    \subsection{Normal Pulsars}
As seen in Fig. \ref{P=1_T6=1_l=2_m=0} and \ref{P=0.1_B12=1_T6=1_l=2_m=0}, the $2\gamma$ mode is not important due to the low temperature in normal pulsars. The properties of the gap was dominated by the CR mode or the resonant ICS mode, especially the CR mode when $K$ is large. More importantly, we find that the CR mode in normal pulsars usually does not meet the $l_e > h_{\rm min}$ criterion, so that $\Delta V$ only increases slowly with increasing $K$, due to decrease of the height. Since particle acceleration is closely related to $\Delta V$, one would expect that the magnetosphere activities would not change noticeably when a significant toroidal oscillation is excited. Observationally there is no noticeable emission property changes associated with glitches in most normal radio pulsars \citep[e.g. for Vela pulsar][]{Helfand2001}. This is understandable within the framework of our model.

Even though the gap properties do not change significantly, glitch-induced oscillations would enlarge the polar cap size, and modify the geometry of the radio beams. \cite{Weltevrede2011} discovered a glitch-induced pulse profile changes in PSR J1119-6127. Normally it has a single-peaked profile. After a large amplitude glitch, it revealed a double-peaked profile for some time, and finally returned to the single-peaked profile. The duration of the double-peaked profile is not well measured, but it is constrained to a range between minutes and months. We suggest that the change of pulse profile is due to the oscillation effect discussed in this paper. After the glitch, oscillations enlarge the polar cap angle and hence, also the size of the radiation cone. Imagine that the line of sight originally cuts across the edge of an emission cone, so that a single peak is observed. After the emission cone is enlarged due to the oscillation effect, the line of sight now cuts through a hollow cone, so that a double-peaked profile is observed. According to the observations of \cite{Weltevrede2011} (their Fig.5), the center of the single-peaked profile is not consistent to that of the double-peaked profile. This is possible within our model, since toroidal oscillations introduce another asymmetry with respect to the misalignment between the magnetic axis and the spin axis. More detailed simulations are needed to testify or falsify this picture. Future radio long-term observations following glitches will bring critical clues to test this model.

    \subsection{AXPs/SGRs}

    Pulsed radio emission is observed after flaring activities of some AXPs/SGRs, which are also accompanied by glitching activities \citep{Kaspi2003,Dib2014,Olausen2014}. This suggests that the condition of radio emission is satisfied after the flares/glitches. In the following, we show that the observations may be understood within the framework of our model.

    The physical nature of AXPs/SGRs is still an open question. The standard model is that they are neutron stars with superstrong magnetic fields, known as magnetars \citep{Thompson1995,Thompson1996}. Alternatively, a solid quark star model was suggested by \cite{Xu2003,Xu2013}. In the following, we discuss the radio activation mechanism within the framework of both models.

\subsubsection{Magnetar model}

    In Fig. \ref{P=10_B12=100_T6=5_l=2_m=0}, we calculate the gap properties for typical parameters of an AXP/SGR:  $P=10\,{\rm s}$, $T=5\times10^6\,{\rm K}$ and $B=10^{14}\,{\rm G}$. Because the magnetic field strength already exceeds the critical value $B_q$, the Klein-Nishina reduction effect makes cross section small, so that $l_e$ becomes very large. The ICS mode would not be important unless we consider the region far from the surface where magnetic field strength is below $B_q$. Since the inner gap height is much shorter than the radius of the pulsar, the resonant ICS mode is not important in any case. The $2\gamma$ mode is more dominant thanks to the high temperature. However, due to strong magnetic field, the CR mode would be more effective than $2\gamma$ mode and would determine the properties of the gap\footnote{It was suggested that photon splitting may suppress pair production in strong magnetic fields \citep{Baring1998,Baring2001}. This requires that both photon modes split. Later studies showed that photon-splitting are only valid in one polarization mode \citep{Usov2002}. The CR and ICS modes discussed in this paper therefore are also relevant in a magnetar environment.}.

    Radio activation of AXPs/SGRs after outburst/glitch may be understood within the following picture: With the strong magnetic field in a magnetar, the CR mode gap would be above the pulsar radio emission deathline even in the quiescent state. However, as slow rotators, the solid angle of the radio emission beam (if exist) would be very small, so that the chance for detecting radio emission from a magnetar during quiescent state is very low. After an outburst/glitch, toroidal oscillations would be excited. The strength of radio emission may be enhanced (due to the larger potential). More importantly, the effect polar cap angle of the pulsar (Eq. (\ref{polar_angle_m=0})) would be increased significantly when oscillations occur. The radio emission beam would be wider, so that it can be detected by an observer on earth. Later, as oscillations damp away, the radio emission decays due to the decrease of the voltage or due to a less favorable geometric configuration. It finally vanishes when the radio beam exits the line of sight. Within this picture, not all outburst/glitches in magnetars would excite radio emission, which seems to be consistent with the data.

    \subsubsection{Solid Quark Star Model}

In this sub-section, we discuss the solid quark star model for AXPs/SGRs \citep{Xu2003,Xu2013}.
Within this model, no strong magnetic field is needed to power outbursts. We therefore assume that AXPs/SGRs have a similar magnetic field as normal pulsars, i.e. $B \sim 10^{12}\,{\rm G}$\footnote{In this model, the spindown behavior of AXPs/SGRs is attributed to the existence of a fossil accretion disk \citep[e.g.][]{Alpar2001}.}. Due to their long periods, these objects are then usually below the radio emission ``death line'', so that they are radio quiet. When an outburst occurs (which attributes to star quakes in the solid quark star model), oscillations are excited. The oscillation-induced potential enhances voltage along the gap, so that one or more gap modes can be activated. The pulsar moves above to the death line, and becomes radio loud. Later, the oscillations are damped out. Radio emission fades away and eventually disappears when the gap state falls below the death line again.

It is worth reviewing here the physical picture of outbursts and toroidal oscillations within the solid quark model.
We consider bulk-variable starquakes \citep{Zhou2014} in solid quark stars. Various quark star equation-of-state defines a general positive $R-M$ relation, with a turn-over at the high $M$ range. There exists a maximum radius $R_{\rm max}$ in the $R-M$ diagram. Due to accretion from a fossil disk, the mass of the quark star increases and might exceed the mass corresponding to $R_{\rm max}$. As mass keeps increasing, the quark star would contract to seek a more stable configuration. Since the star is in the solid state, the elastic energy would be accumulated to resist radius contraction.
When the accumulated elastic energy finally exceeds a threshold, the solid would crack, and a bulk-variable starquake would happen. Strong oscillations are then excited. The gravitational and elastic energy released in a starquake may be estimated as
\begin{equation}\label{energy}
    \frac{GM^2}{R}\frac{\Delta R}{R} \sim 10^{53}\frac{\Delta R}{R}\,{\rm erg}
\end{equation}
for typical parameters: $M\sim 1.4M_\odot$, $R\sim10^6\,{\rm cm}$. Here $\Delta R$ is change of radius during configuration rearrangement. Such an energy is certainly large enough to power AXP/SGR outbursts and even giant flares. It has been suggested that oscillation-driven radiation could also act as the radiation mechanism of giant flares within this model \citep{Xu2006}. The starquakes induce oscillations, which convert the gravitational and elastic energy to the mechanical energy of oscillations. The oscillations then lead to acceleration of particles based on the mechanisms discussed in this paper, so that the mechanical energy of oscillations is converted to the kinetic energy of the particles, and is finally converted to high energy photon energy via CR or ICS mechanisms. An SGR/AXP outburst is then produced.
The kHz quasi-periodic oscillations (QPOs) observed after SGR giant flares \citep[e.g.][]{Israel2005} may hint towards global oscillations of the central compact star, which is consistent with this picture.

\begin{figure*}
    \centering
    \includegraphics [width=15cm]{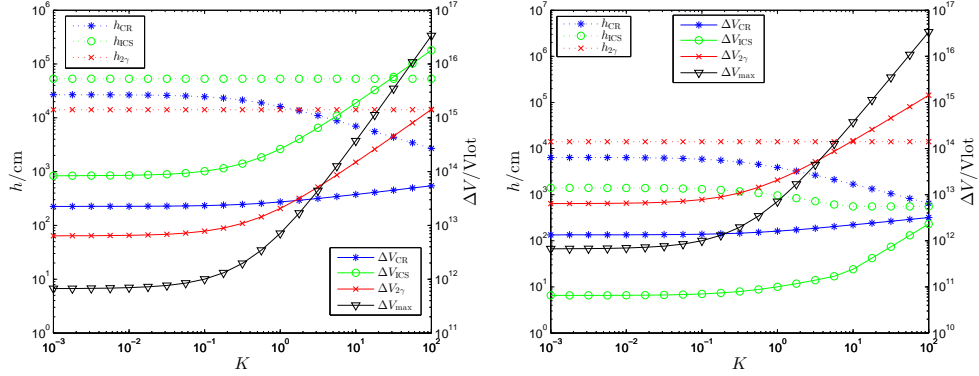}
    \caption{\label{P=2.07_B12=1_T6=5_l=2_m=0} Gap properties of 1E 1547.0-5408 as functions of dimensionless oscillations amplitude $K$. A dipole magnetic field $B=10^{12}\,{\rm G}$ is assumed. Other parameters are $P=2.07\,{\rm s}$, $T=5\times10^6\,{\rm K}$. The black solid line with inverted triangle markers represent the maximum voltage across the polar cap. If $\Delta V$ is less than this critical value, the gap becomes above the death line, and radio emission is reactivated. Other lines share the same meaning with Fig. \ref{P=1_T6=1_l=2_m=0}. The left and right panels are for the dipole and multipole magnetic field configurations, respectively.}
\end{figure*}

    Along the similar line, below we show that transient pulsed radio emission of several AXPs/SGRs detected after outbursts or giant flares is also consistent with such a picture.
    We calculate the gap properties of the four radio-loud AXPs/SGRs with different oscillation amplitudes. The magnetic field is assumed as $10^{12}\,{\rm G}$ in order to explain lack of radio emission in the quiescent state (below the death line). As an example, we show the result of 1E 1547.0-5408 in Fig. \ref{P=2.07_B12=1_T6=5_l=2_m=0}. The black solid line with inverted triangle represent the maximum voltage across the polar cap, i.e. $\Delta V$ less than this critical value is necessary for radio emission. Notice that the results of $K=10^{-3}$ can represent the gap properties without oscillation.
    We can see that with a dipole magnetic field consistent, the result is consistent with our expectation: In the quiescent state, there is no oscillation, $\Delta V$ of all three mode are larger than $\Delta V_{\rm max}$, so that the star is radio quiet. After an outburst, stellar oscillations are excited. As long as $K$ exceeds a critical value, there is at least one mode whose $\Delta V$ becomes less than $\Delta V_{\rm max}$, so that the gap is activated and the star would become radio loud. With oscillations damped away later, radio emission decays and finally vanishes.

    We note that a dipole magnetic field is not a necessary condition.
For a multipole field, gaps are more likely above the death line. However, if the magnetic field strength is even weaker (compared with the typical value $10^{12}\,{\rm G}$), the gap would be below the death line in the quiescent state. The mechanism discussed above then becomes valid.

    For other AXPs/SGRs with radio activation, we also carried out corresponding calculations. Similar to the result of 1E 1547.0-5408, we find that in a dipolar magnetic field with $B=10^{12}\,{\rm G}$, there always exists a critical $K$ value.
    When $K$ exceeds the critical value, the gaps would be activated above the death line. In Table \ref{T 1}, we show the critical values of $K$ of the four radio AXPs/SGRs.
    For solid quark stars, a relatively large $K$ would be possible. This may be justified by the following two arguments \citep[e.g.][]{Xu2006}. First, in solid quark stars, since the entire star is a rigid solid body, the large gravitational and elastic energy released in star quakes (Eq. \ref{energy}) are large enough to power the large-amplitude oscillations required by the model. For neutron stars, only the crust is believed to be solid. In order to attain the large oscillation energy, one must appeal to the magnetic energy reservoir of the magnetar. Second, since solid quark stars are self-bound via strong interaction, unlike the gravity-bound case for neutron stars, the surface velocity in solid quark stars can be close to or even exceed the Kepler velocity. This allows the star to reach a $K$ value greater than unity.
    We also note that here we only calculate the $l=2, m=0$ mode. For a larger $l$, the critical value of $K$ will decrease markedly, which would ease the radio reactivation condition. Our results suggest that all the radio-activating AXPs/SGRs can be well interpreted in a reasonable parameter space within this model.

\begin{table}
    \begin{center}
        \begin{tabular}{cccc}
            \hline
            Radio AXPs/SGRs & $P/{\rm s}$ & $T/{\rm 10^6K}$ & critical $K$ \\
            \hline
            1E 1547.0-5408 & 2.07 & 5.0  & $2.54$ (CR) \\
            PSR J1622-4950 & 4.33 & 5.8  & $3.20$ ($2\gamma$) \\
            XTE J1810-197 & 5.54 & 2.1  & $8.90$ (CR) \\
            SGR J1745-2900 & 3.76 & 5.0\tablenotemark{1} & $5.53$ (CR) \\
            \hline
        \end{tabular}
        \tablenotetext{1}{Since the surface temperature of the SGR J174-2900 is unknown, we just adopt a typical value.}
        \caption{\label{T 1}The critical $K$ of the four AXPs/SGRs that show radio reactivation. When $K$ exceeds this value, $\Delta V$ will become less than $\Delta V_{\rm max}$, and radio emission would be activated. A dipole field with $B=10^{12}\,{\rm G}$ is assumed, and the $l$=2, $m$=0 oscillation mode is calculated.}
    \end{center}
\end{table}

\section{CONCLUSIONS AND DISCUSSIONS}
Along with rotation itself, toroidal stellar oscillations can provide additional voltage due to unipolar induction, change Goldreich-Julian charge density, and influence particle acceleration and properties of the pulsar inner gap. In this paper, we studied in detail the gap properties under toroidal oscillations for three different modes: CR, ICS, and $2\gamma$. We show that as the oscillation amplitude varies, inner gap mode changing could happen, including activating a radio pulsar below the radio emission death line. Oscillations can also enlarge the size of the effective polar cap. These interesting properties allow us to interpret the observed radio activation in several AXPs/SGRs within the frameworks of both magnetar model and solid quark star model. Within the magnetar model, the reactivation is interpreted as a geometric effect due to the enlargement of the effective polar cap size. Within the solid quark star model, on the other hand, the reactivation is interpreted as moving a low $B$ star from below the death line to above via oscillations.  The same model can also account for the post-glitch radio pulse profile change observed in PSR J1119-6127.

As discussed earlier in the paper, we only consider the maximum effect of stellar oscillations and ignore the periodical variations of the gap properties. The time scale for periodic variation is too short to give interesting observational effects, even though within a period, gap mode changing may still happen. Also when threshold condition of radio emission is considered (death line), the maximum effect would be the most relevant one to define the threshold condition.

Since the energy of oscillation is proportional to the square of amplitude, $K=\tilde{\eta}/\Omega B$ denotes the ratio of the oscillation velocity and the rotational velocity. The total energy of oscillation may be expressed as $E\propto K^2$. We may estimate it as $K^2$ times of the rotational energy, i.e.
\begin{equation}
    E \sim \frac{1}{2} I \Omega^2 K^2 = 1.58\times10^{46} M_1 R_6^2 P^{-2} K^2 \,{\rm erg},
\end{equation}
where $M_1$ and $R_6$ are the mass and radius of the pulsar normalized to $1 M_\odot$ and $10^6$ cm, respectively, and $P$ is the period of the pulsar in units of second.
Within this model, the oscillation energy should be larger than the observed energy release. This can set a rough lower limit to $K$.

In order to have an observational interest, a relatively large $K$ is needed. This is naturally expected within the solid quark star model, since oscillation occur in the entire star and the oscillation amplitude (associated with $K$) could be very large due to the solid rigidity. For neutron stars, on the other hand, oscillations may only occur in the crust. A large amplitude oscillation may break the crust. On the other hand, we note again that for oscillation modes with a higher $l$ value, a relatively smaller oscillation velocity would be needed in order have a same oscillation effect, as shown in  Eq. (\ref{rho_GJ_ratio}).
Furthermore, the oscillation velocities can vary largely at different locations on the neutron (quark) star's surface. For rotation, on the other hand, different points on the star's surface would have the same anglular velocity $\bm{\Omega}$, and $\rho_{\rm GJ,rot}\propto\bm{\Omega}\cdot\bm{B}$; but for oscillations, different points on the star's surface would have different anglular velocity component perpendicular to the direction of the magnetic field, denoted as $\Omega_{\rm eff\perp}$, and $\rho_{\rm GJ,osc}$ depends on $\Omega_{\rm eff\perp}$ in the polar cap region. For some oscillation modes, $\Omega_{\rm eff\perp}$ could be small in the equatorial region while large in the polar cap region. Therefore significant oscillation effects could be obtained without breaking the crust of a neutron star.

One may ask why not all AXPs/SGRs have radio emission after outbursts. Within the magnetar model, this may be interpreted as a viewing effect: not all activated radio emission can enter our line of sight. Within the solid quark star model, besides beaming effect, there is another possibility. Radio activation is related moving a pulsar from the grave yard, however, given certain parameters, even after oscillations the pulsar is still below the death line. So radio emission would not be observed in any case.

The solid quark star model requires a fossil disk to spin down the star. One related question would be whether the disk would absorb the radio emission from the pulsar magnetosphere. The fossil disks are believed to be very thin, so the probability of radio emission being absorbed should not be very large. In any case, when significant absorptions happen, this model would not produce radio emission in any case. These may correspond to those AXPs/SGRs that never activate.

\paragraph{Acknowledgments}

We thank an anonymous referee for a constructive report, Vicky Kaspi for drawing our attention to the observations of PSR J1119-6127, and Viktoriya Morozova for a helpful communication.
MXL and RXX would like to thank useful discussions at the pulsar group meetings at Peking University.
This work is supported by the National Basic Research Program of China (973 program) No. 2012CB821801,
the National Natural Science Foundation of China (Grant Nos. 11225314, Grant Nos. 11133002, NSFC-11103020),
National Found for Fostering Talents of Basic Science (No. J0630311)
and the Undergraduate Research Fund of Education Foundation of Peking University.

\appendix
\section{VOLTAGE ACROSS POLAR CAP AND THE POLAR CAP ANGLE}
For simplification, we assume that the magnetic axis aligns with the spin axis, so that $\vartheta=\theta$.

To obtain the voltage across the polar cap $\Delta V_{\rm max}$, we have to do the integration
\begin{equation} \label{V_max_int}
    \Delta V_{\rm max} = \int_0^{\Theta_0} E^\theta(R,\theta,\phi)d\theta.
\end{equation}
Due to unipolar induction, we have
\begin{equation}
    E = -\frac{\bm{v}\times\bm{B}}{Nc},
\end{equation}
where $\bm{v}$ includes contributions from both rotation and oscillation. The magnetic field takes the form
\begin{equation}
    B^r = B_0\frac{f(\tilde{r})}{f(1)}\tilde{r}^3\cos\theta,\
    B^\theta = \frac{1}{2}B_0N \left[-2\frac{f(\tilde{r})}{f(1)}+\frac{3}{(1-\varepsilon/\tilde{r})f(1)}\right] \frac{\sin\theta}{\tilde{r}^3}.
\end{equation}
With Eq. (\ref{oscillation}), we get
\begin{equation} \label{E}
    E^\theta(R,\theta,\phi) = -\frac{1}{Nc} \left[\Omega R \tilde{r}\sin\theta(1-\frac{\kappa}{\tilde{r}^3})-\partial_\theta Y_{lm} \tilde{\eta}(\tilde{r})\right] B_0\frac{f(\tilde{r})}{f(1)} \frac{\cos\theta}{\tilde{r}^3}.
\end{equation}
Substitute Eq. (\ref{E}) into Eq. (\ref{V_max_int}), we have
\begin{equation}
    \Delta V_{\rm max} = -\frac{\Omega R^2NB_0}{c} \left[(1-\kappa)\frac{\sin^2\Theta_0}{2} - K\int_0^{\Theta_0}\partial_\theta Y_{lm}(\theta,\phi)\cos\theta d\theta\right].
\end{equation}
For a small polar angle $\Theta_0 \ll 1$, approximately one has
\begin{equation}
    \int_0^{\Theta_0}\partial_\theta Y_{lm}(\theta,\phi)\cos\theta d\theta \approx -\frac{l(l+1)}{2}Y_{l0}\frac{\Theta^2}{2},\ {\rm for}\ m=0;
\end{equation}
\begin{equation}
    \int_0^{\Theta_0}\partial_\theta Y_{lm}(\theta,\phi)\cos\theta d\theta \approx Y_{lm}(\Theta_0,\phi),\ {\rm for}\ m\neq0.
\end{equation}
We then obtain the expression of $\Delta V_{\rm max}$, see Eqs. (\ref{V_max_m=0}) and (\ref{V_max_m!=0}).

For an oscillating star, the last open field line is not straightforwardly defined. One may define a last open field line as the field line whose intersection with the equitorial plane satisfies the condition that the kinetic energy density of the outflowing plasma equals to the energy density of the magnetic field \citep[e.g.]{Abdikamalov2009,Morozova2010}:
\begin{equation}
    \epsilon_{pl}(R_a,\pi/2,\phi) = \epsilon_{em}(R_a,\pi/2,\phi),
\end{equation}
where $R_a$ denote the radial coordinate of the intersection point at the equator, $\epsilon_{pl}$ and $\epsilon_{em}$ are the kinetic energy density and the magnetic field density, respectively, which take the form \citep{Abdikamalov2009}
\begin{equation}
    \epsilon_{pl}(R_a,\pi/2,\phi) = \frac{1}{2f(1)} \frac{N_R}{N_{R_a}} \frac{R^3}{R_a^3} \Delta V_{\rm max} \rho_{\rm GJ},
\end{equation}
\begin{equation}
    \epsilon_{em}(R_a,\pi/2,\phi) = \frac{N_{R_a}}{32\pi}\frac{R^6}{R_a^6}B_0^2.
\end{equation}
Here $N_R=\sqrt{1-2M/R}$ and $N_{R_a}=\sqrt{1-2M/R_a}\approx1$.
The footprints of these field lines in the polar region at surface define the polar cap.
So the polar cap angle can be defined through the relation
\begin{equation}
    \frac{R}{R_a} = \frac{f(1)}{f(R_a/R)}\Theta_0^2.
\end{equation}
Here the function $f(x)$ is the correction factor for general relativity, which is defined in Eq.(\ref{f}). Since $R_a \gg R$, one has $f(R_a/R)\approx1$ . Substitute Eqs. (\ref{rho_GJ_rot}), (\ref{rho_GJ_osc}), (\ref{V_max_m=0}), (\ref{V_max_m!=0}) into the above equations, we can derive the expression of the polar cap angle, which are Eqs. (\ref{polar_angle_m=0}) and (\ref{polar_angle_m!=0}) in the text.
\citet{Morozova2010} have carried out a similar derivation, but their results are somewhat different from ours.


\begin{thebibliography}{}
\expandafter\ifx\csname natexlab\endcsname\relax\def\natexlab#1{#1}\fi

\bibitem[{{Abdikamalov} {et~al.}(2009){Abdikamalov}, {Ahmedov}, \&
  {Miller}}]{Abdikamalov2009}
{Abdikamalov}, E.~B., {Ahmedov}, B.~J., \& {Miller}, J.~C. 2009, \mnras, 395,
  443

\bibitem[{{Alpar}(2001)}]{Alpar2001}
{Alpar}, M.~A. 2001, \apj, 554, 1245

\bibitem[{{Baring} \& {Harding}(1998)}]{Baring1998}
{Baring}, M.~G., \& {Harding}, A.~K. 1998, \apjl, 507, L55

\bibitem[{{Baring} \& {Harding}(2001)}]{Baring2001}
---. 2001, \apj, 547, 929

\bibitem[{{Baring} {et~al.}(2011){Baring}, {Wadiasingh}, \&
  {Gonthier}}]{Baring2011}
{Baring}, M.~G., {Wadiasingh}, Z., \& {Gonthier}, P.~L. 2011, \apj, 733, 61

\bibitem[{{Cameron} {et~al.}(2005){Cameron}, {Chandra}, {Ray}, {Kulkarni},
  {Frail}, {Wieringa}, {Nakar}, {Phinney}, {Miyazaki}, {Tsuboi}, {Okumura},
  {Kawai}, {Menten}, \& {Bertoldi}}]{Cameron2005}
{Cameron}, P.~B., {Chandra}, P., {Ray}, A., {et~al.} 2005, \nat, 434, 1112

\bibitem[{{Dermer}(1990)}]{Dermer1990}
{Dermer}, C.~D. 1990, \apj, 360, 197

\bibitem[{{Dib} \& {Kaspi}(2014)}]{Dib2014}
{Dib}, R., \& {Kaspi}, V.~M. 2014, \apj, 784, 37

\bibitem[{{Frail} {et~al.}(1999){Frail}, {Kulkarni}, \& {Bloom}}]{Frail1999}
{Frail}, D.~A., {Kulkarni}, S.~R., \& {Bloom}, J.~S. 1999, \nat, 398, 127

\bibitem[{{Gaensler} {et~al.}(2005){Gaensler}, {Kouveliotou}, {Gelfand},
  {Taylor}, {Eichler}, {Wijers}, {Granot}, {Ramirez-Ruiz}, {Lyubarsky},
  {Hunstead}, {Campbell-Wilson}, {van der Horst}, {McLaughlin}, {Fender},
  {Garrett}, {Newton-McGee}, {Palmer}, {Gehrels}, \& {Woods}}]{Gaensler2005}
{Gaensler}, B.~M., {Kouveliotou}, C., {Gelfand}, J.~D., {et~al.} 2005, \nat,
  434, 1104

\bibitem[{{Gil} {et~al.}(2006){Gil}, {Melikidze}, \& {Zhang}}]{Gil2006}
{Gil}, J., {Melikidze}, G., \& {Zhang}, B. 2006, \apj, 650, 1048

\bibitem[{{Gil} {et~al.}(2003){Gil}, {Melikidze}, \& {Geppert}}]{Gil2003}
{Gil}, J., {Melikidze}, G.~I., \& {Geppert}, U. 2003, \aap, 407, 315

\bibitem[{{Goldreich} \& {Julian}(1969)}]{Goldreich1969}
{Goldreich}, P., \& {Julian}, W.~H. 1969, \apj, 157, 869

\bibitem[{{Haensel} {et~al.}(2007){Haensel}, {Potekhin}, \&
  {Yakovlev}}]{Haensel2007}
{Haensel}, P., {Potekhin}, A.~Y., \& {Yakovlev}, D.~G., eds. 2007, Astrophysics
  and Space Science Library, Vol. 326, {Neutron Stars 1 : Equation of State and
  Structure}

\bibitem[{{Helfand} {et~al.}(2001){Helfand}, {Gotthelf}, \&
  {Halpern}}]{Helfand2001}
{Helfand}, D.~J., {Gotthelf}, E.~V., \& {Halpern}, J.~P. 2001, \apj, 556, 380

\bibitem[{{Israel} {et~al.}(2005){Israel}, {Belloni}, {Stella}, {Rephaeli},
  {Gruber}, {Casella}, {Dall'Osso}, {Rea}, {Persic}, \&
  {Rothschild}}]{Israel2005}
{Israel}, G.~L., {Belloni}, T., {Stella}, L., {et~al.} 2005, \apjl, 628, L53

\bibitem[{{Kaspi} {et~al.}(2003){Kaspi}, {Gavriil}, {Woods}, {Jensen},
  {Roberts}, \& {Chakrabarty}}]{Kaspi2003}
{Kaspi}, V.~M., {Gavriil}, F.~P., {Woods}, P.~M., {et~al.} 2003, \apjl, 588,
  L93

\bibitem[{{Medin} \& {Lai}(2007)}]{Medin2007}
{Medin}, Z., \& {Lai}, D. 2007, \mnras, 382, 1833

\bibitem[{{Mereghetti}(2008)}]{Mereghetti2008}
{Mereghetti}, S. 2008, \aapr, 15, 225

\bibitem[{{Morozova} {et~al.}(2010){Morozova}, {Ahmedov}, \&
  {Zanotti}}]{Morozova2010}
{Morozova}, V.~S., {Ahmedov}, B.~J., \& {Zanotti}, O. 2010, \mnras, 408, 490

\bibitem[{{Morozova} {et~al.}(2012){Morozova}, {Ahmedov}, \&
  {Zanotti}}]{Morozova2012}
---. 2012, \mnras, 419, 2147

\bibitem[{{Muslimov} \& {Tsygan}(1992)}]{Muslimov1992}
{Muslimov}, A.~G., \& {Tsygan}, A.~I. 1992, \mnras, 255, 61

\bibitem[{{Olausen} \& {Kaspi}(2014)}]{Olausen2014}
{Olausen}, S.~A., \& {Kaspi}, V.~M. 2014, \apjs, 212, 6

\bibitem[{{Ruderman} \& {Sutherland}(1975)}]{Ruderman1975}
{Ruderman}, M.~A., \& {Sutherland}, P.~G. 1975, \apj, 196, 51

\bibitem[{{Thompson} \& {Duncan}(1995)}]{Thompson1995}
{Thompson}, C., \& {Duncan}, R.~C. 1995, \mnras, 275, 255

\bibitem[{{Thompson} \& {Duncan}(1996)}]{Thompson1996}
---. 1996, \apj, 473, 322

\bibitem[{{Unno} {et~al.}(1989){Unno}, {Osaki}, {Ando}, {Saio}, \&
  {Shibahashi}}]{Unno1989}
{Unno}, W., {Osaki}, Y., {Ando}, H., {Saio}, H., \& {Shibahashi}, H. 1989,
  {Nonradial oscillations of stars} (Tokyo: Univ. Tokyo Press)

\bibitem[{{Usov}(2002)}]{Usov2002}
{Usov}, V.~V. 2002, \apjl, 572, L87

\bibitem[{{Wang} {et~al.}(2005){Wang}, {Wu}, {Fan}, {Dai}, \&
  {Zhang}}]{Wang2005}
{Wang}, X.~Y., {Wu}, X.~F., {Fan}, Y.~Z., {Dai}, Z.~G., \& {Zhang}, B. 2005,
  \apjl, 623, L29

\bibitem[{{Weltevrede} {et~al.}(2011){Weltevrede}, {Johnston}, \&
  {Espinoza}}]{Weltevrede2011}
{Weltevrede}, P., {Johnston}, S., \& {Espinoza}, C.~M. 2011, \mnras, 411, 1917

\bibitem[{Xia {et~al.}(1985)Xia, Qiao, Wu, \& Hou}]{Xia1985}
Xia, X., Qiao, G., Wu, X., \& Hou, Y. 1985, \aap, 152, 93

\bibitem[{{Xu}(2013)}]{Xu2013}
{Xu}, R. 2013, Scientia Sinica Physica, Mechanica and Astronomica, 43, 1288

\bibitem[{{Xu}(2003)}]{Xu2003}
{Xu}, R.~X. 2003, \apjl, 596, L59

\bibitem[{{Xu} {et~al.}(1999){Xu}, {Qiao}, \& {Zhang}}]{Xu1999}
{Xu}, R.~X., {Qiao}, G.~J., \& {Zhang}, B. 1999, \apjl, 522, L109

\bibitem[{{Xu} {et~al.}(2006){Xu}, {Tao}, \& {Yang}}]{Xu2006}
{Xu}, R.~X., {Tao}, D.~J., \& {Yang}, Y. 2006, \mnras, 373, L85

\bibitem[{{Zanotti} {et~al.}(2012){Zanotti}, {Morozova}, \&
  {Ahmedov}}]{Zanotti2012}
{Zanotti}, O., {Morozova}, V., \& {Ahmedov}, B. 2012, \aap, 540, A126

\bibitem[{{Zhang}(2001)}]{Zhang2001}
{Zhang}, B. 2001, \apjl, 562, L59

\bibitem[{{Zhang} {et~al.}(2000){Zhang}, {Harding}, \& {Muslimov}}]{Zhang2000}
{Zhang}, B., {Harding}, A.~K., \& {Muslimov}, A.~G. 2000, \apjl, 531, L135

\bibitem[{Zhang \& Qiao(1996)}]{Zhang1996}
Zhang, B., \& Qiao, G. 1996, \aap, 310, 135

\bibitem[{Zhang \& Qiao(1998)}]{Zhang1998}
Zhang, B., \& Qiao, G.~J. 1998, \aap, 338, 62

\bibitem[{{Zhang} {et~al.}(1997){Zhang}, {Qiao}, {Lin}, \& {Han}}]{Zhang1997}
{Zhang}, B., {Qiao}, G.~J., {Lin}, W.~P., \& {Han}, J.~L. 1997, \apj, 478, 313

\bibitem[{{Zhou} {et~al.}(2014){Zhou}, {Lu}, {Tong}, \& {Xu}}]{Zhou2014}
{Zhou}, E.~P., {Lu}, J.~G., {Tong}, H., \& {Xu}, R.~X. 2014, \mnras, 443, 2705

\end{thebibliography}
\end{document}